# Designing materials for plasmonic systems

Martin G. Blaber, Matthew D. Arnold, Michael J. Ford*
*Institute for Nanoscale Technology, Dept of Physics and Advanced Materials, University of Technology, Sydney.*
*Corresponding Author: Mike.Ford@uts.edu.au

**Abstract**
We use electronic structure calculations based upon density functional theory to search for ideal plasmonic materials among the alkali noble intermetallics. Importantly, we use density functional perturbation theory to calculate the electron-phonon interaction and from there use a first order solution to the Boltzmann equation to estimate the phenomenological damping frequency in the Drude dielectric function. We discuss the necessary electronic features of a plasmonic material and investigate the optical properties of the alkali-noble intermetallics in terms of some generic plasmonic system quality factors. We conclude that at low negative permittivities, KAu with a damping frequency of 0.0224 eV and a high optical gap to bare plasma frequency ratio, outperforms gold and to some extent silver as a plasmonic material. Unfortunately, a low plasma frequency (1.54 eV) reduces its utility in modern plasmonics applications. We also discuss, briefly, the effect of local fields on the optical properties of these materials.

## I. INTRODUCTION

Plasmonic systems are finding a variety of uses in negative index materials[1, 2] and superlensing [3, 4], thermal therapeutics[5, 6] and waveguiding [7] to name a few. The idea of designing new materials for these plasmonics applications is appealing. Gold and silver are exclusively used mainly because of experimental convenience. Even a cursory investigation using measured optical constants for the elements reveals far superior alternatives, at least as far as optical metrics are concerned [8]. The alkali metals, although ideal in this regard, are naturally quite difficult to work with. The next obvious place to search is the metallic alloys and compounds where a suitable compromise between metric performance and reactivity may be possible. The choice now becomes bewildering and disappointingly few experimental compilations of optical constants are available. First principles calculations of the optical response offer a promising way forward. We have already reported such studies where KAu was identified as a promising candidate [9]. These calculations were performed within the framework of Density Functional Theory, using the Random Phase Approximation to calculate the optical response. However, they provide no phenomenological electron scattering term, and hence are only capable of giving an accurate description of the interband component of the dielectric response and position of the plasma frequency. Assuming a Drude intraband component and using an empirical value for the damping, the imaginary component of the dielectric function can be constructed. As we have also pointed out [8] the real part is also required to assess optical performance. Although this is straightforward to calculate from a Kramers-Kronig analysis, knowledge of the imaginary component over a broad frequency range is required to achieve reliable results. This requires both knowledge of the plasma frequency and phenomenological Drude damping. In this paper we address this shortcoming by calculating the electron-phonon coupling for some of the metallic compounds studied in [9]. This allows us to derive the DC resistivity and hence gain a reasonable estimate of the damping term.

The fundamental problem with plasmonic systems is metallic loss. Although a variety of work has been done to introduce gain into these systems[10-13], within reasonable bounds a passive system has the advantage of simplicity. The specifics of designing a geometry that performs best in a particular plasmonic application is non trivial, and in real systems there are a number of loss mechanisms including surface scattering of electrons on the boundaries of the geometry[14, 15], surface states (see for example [16, 17]), surface roughness and grain boundaries[18]. Some of these effects can be minimized by reverse engineering of the geometry[19].

We have previously assessed the performance of various elements as plasmonic materials, with specific examples including nanospheres[20], the so called 'poor mans' superlens [21] and nanoshells[14]. More recently, we presented an analysis of the plasmonic performance of various elements using a set of geometry specific metrics[8]. A convenient way to describe the operation of a plasmonic system is in terms of the real part of the dielectric function $\epsilon'$, as this is directly related to the geometry in question. The free electron metals Al, Ag, Au, Na and K outperform most other elements, each in their own frequency and permittivity range. We shall describe from hereon systems with low negative permittivity as having low permittivity. For high frequencies/low permittivities, Al performs most favorably, while sodium and potassium perform very well at medium permittivities $-\epsilon' < 50$ and gold performs favorably at high permittivities $-\epsilon' > 100$ [8].

The number of free-electron-like metallic elements in the periodic table severely limits the possibilities for passive plasmonic systems, and as such an alternative must be realized[9].

A number of alloy systems have been used in plasmonics, the optical properties of random mixtures of Au, Ag, Cu and Al were studied by Sharma et al [22]. They treat the bulk dielectric function as a linear combination of the free electron dielectric functions of the constituent metals:

$$\epsilon_{\text{alloy}}(\omega) = x\epsilon_{\text{metal}_1}(\omega) + (1-x)\epsilon_{\text{metal}_2}(\omega) \qquad (1)$$

They employ a Drude model of the form

$$\epsilon(\omega) = 1 - \frac{\omega_p^2}{\omega(\omega + i\gamma)}, \quad (2)$$

where $\omega_p$ and $\gamma$ are the plasma frequency and phenomenological damping terms respectively. They suggest that the sensitivity and signal-to-noise ratio (SNR) of a surface plasmon based fiber optic sensor can be tuned, with Al having the best SNR and gold the best sensitivity. Unfortunately, single electron excitations due to clustering or stoichiometric alloying (such as $AuAl_2$ [23]) are present in all these alloys, and can have a dramatic effect on the quality of a surface plasmon resonance.

Theoretical [24] and experimental [25] work on SiC inclusions in $MgB_2$ has suggested that because of low $\gamma$ in $MgB_2$ compared to silver and gold, an isotropic negative index material is possible in the visible regime. Bobb et al [26] have recently demonstrated a shifting of the optical gap (frequency of the onset of interband transitions) with the introduction of small percentages of Cd into Au, leading to a slight increase in the absorption efficiency of nanospheres compared to those made of elemental gold. Unfortunately, the alloying disrupts the near infrared behavior of gold where its local surface plasmon performance is optimal [27].

Here we shall discuss the optimum electronic features for an ideal plasmonic material, and apply these to some Alkali-Noble intermetallics. We have previously discussed [9] the optical properties of the noble-group-III (NG), alkali-noble-group-III (ANG) and alkali-noble (AN) intermetallics in terms of the interband contribution to the dielectric function. The alkali-noble intermetallics showed great promise with KAu having a band edge above the plasma frequency, indicating that it should perform very well at low permittivities. However, in that work we neglected local field effects and lacked an effective method for calculating the Drude phenomenological damping term, so we could not adequately assess the absolute plasmonic performance of these compounds. Here, to complete our analysis, we have calculated the interband contribution to the dielectric function including local field effects within the Random Phase Approximation (RPA) and calculated the intraband contribution by approximating the phenomenological damping term from first principles using the phonon limited DC resistivity. The real part of the permittivity is provided by a Kramers-Kronig integration.

## II. THE IDEAL PLASMONIC MATERIAL

There are four main electronic features necessary for a good plasmonic material. (i) The gradient of bands at the Fermi surface must be high enough to allow for an appropriate plasma frequency $\omega_p$. (ii) The Drude phenomenological damping term $\gamma$, must be low compared to the plasma frequency. (iii) The "core polarizability", represented in the Drude model by $\epsilon_\infty$ must be low. And finally, (iv) the ratio between the optical gap, $\omega_g$, and the plasma frequency must be proportional to the sharpness of the band edge ($\epsilon''(\omega_g)$) itself. i.e. a material with a sharp band edge that reduces rapidly with increased frequency will perform better than a material where the band edge does not comprise the dominant transition mechanism. This amounts to the screened bulk plasmon frequency, $\omega_s$, being well separated from the optical gap. As an example, in gold, interband transitions interrupt the plasmonic response of thin films and nanospheres because of low $\omega_s/\omega_g$. In silver, where the ratio is slightly higher, the plasmonic performance increases substantially.

The complex permittivity $\epsilon(\omega) = \epsilon' + i\epsilon''$ in the Drude approximation is given by

$$\epsilon(\omega) = \epsilon_\infty - \frac{\omega_p^2}{\omega(\omega + i\gamma)} \quad (3)$$

There are an infinite number of metrics derivable for the quality of a plasmonic resonance in any conceivable plasmonic system, here we shall discuss two generic, system quality factors. Within the quasistatic approximation, where the system features are substantially smaller than the wavelength, and in the limit of low loss ($\epsilon''$), a generic metric for the quality of a localized surface plasmon takes the form:

$$Q_{LSP} = -\epsilon'/\epsilon''. \quad (4)$$

Applicable examples include transverse prolate ellipsoid (nanorod) modes, oblate ellipsoid film modes, nanospheres and nanoshells (assuming the shell is not so thin that surface scattering dominates), see for example [14].

The derivative of $Q_{LSP}$ with respect to frequency, gives a good indication of the quality of the long wavelength behavior of local surface plasmon modes. In a Drude metal the frequency which gives maximum $Q_{LSP}$ is:

$$\omega_{max}^{Q_{LSP}} = \left(\frac{\omega_p^2 - \gamma \epsilon_\infty}{3\epsilon_\infty}\right)^{1/2}, \quad (5)$$

and the associated quality factor

$$Q_{LSP}^{max}(\omega_{max}^{Q_{LSP}}) = \frac{2(\omega_p^2 - \gamma^2 \epsilon_\infty)^{3/2}}{3\gamma \omega_p^2 \sqrt{3\epsilon_\infty}} \quad (6)$$

For less localized modes, such as the longitudinal mode on a prolate ellipse, or traveling waves on a metal-dielectric interface, a reasonably generic quality factor, which is proportional to the propagation length of an SPP at a metal dielectric interface in the limit of low loss can be written [28]:

$$Q_{SPP} = \epsilon'^2/\epsilon''. \quad (7)$$

In a Drude model, the SPP resonance quality at a given permittivity can be written:

$$Q_{SPP} = \frac{\varepsilon'^2 \sqrt{\gamma^2(\epsilon_\infty - \epsilon') - \omega_p^2}}{\gamma(\epsilon' - \epsilon_\infty)^{3/2}} \quad (8)$$

To a good approximation ($\epsilon_\infty = 0$) $Q_{SPP}$ increases as $\omega \to \gamma$, with a maximum value:

$$Q_{SPP}^{max} = \frac{\omega_p^2}{2\gamma^2} \qquad (9)$$

The temperature dependent phenomenological damping term $\gamma$ can be approximated from the phonon limited DC resistivity by[29]:

$$\gamma(T) = \epsilon_0 \omega_p^2 \rho_{DC}(T), \qquad (10)$$

where $\epsilon_0$ is the permittivity of free space and we use a temperature of 300K for the rest of this paper.

## III. METHOD

We have calculated the ground state, optical and dynamical properties of the alkali-noble intermetallics using Density Functional Theory (DFT). As we shall only be discussing low energy photons, and bulk materials, we shall take the $\mathbf{q} \to \mathbf{0}$ limit. The macroscopic dielectric function, including local field effects, can be written in terms of the $(\mathbf{G},\mathbf{G}')=(0,0)$ element of the inverse microscopic dielectric function:

$$\epsilon(\omega) = \lim_{\mathbf{q} \to \mathbf{0}} \frac{1}{[\epsilon(\mathbf{q},\omega)^{-1}]_{\mathbf{G}=0,\mathbf{G}'=0}} \qquad (11)$$

Here, only a small number of reciprocal lattice vectors $\mathbf{G}$ are required for convergence. Within the RPA, the microscopic dielectric function can be written in terms of the non-interacting response function $\chi$ [30-32]:

$$\epsilon_{\mathbf{G},\mathbf{G}'}(\mathbf{q},\omega) = \delta_{\mathbf{G},\mathbf{G}'} - \frac{4\pi e^2}{|\mathbf{q}+\mathbf{G}|^2}[\chi(\mathbf{q},\omega)]_{\mathbf{G},\mathbf{G}'}, \qquad (12)$$

where the linear response function is given by:

$$[\chi(\mathbf{q},\omega)]_{\mathbf{G},\mathbf{G}'} = -\frac{2}{\Omega}\sum_{i,j,\mathbf{k}} \frac{f_{\mathbf{k}-\mathbf{q}i}(1-f_{\mathbf{k}j})}{\omega - \varepsilon_{j\mathbf{k}} + \varepsilon_{i\mathbf{k}-\mathbf{q}} + i\delta} \\ \times p_{ij\mathbf{k}}^*(\mathbf{q},\mathbf{G}) p_{ij\mathbf{k}}(\mathbf{q},\mathbf{G}') \qquad (13)$$

Here, $\Omega$ is the Brillouin zone volume, $f$ the occupation, $p_{ij\mathbf{k}}(\mathbf{q},\mathbf{G})$ the matrix elements given by:

$$p_{ij\mathbf{k}}(\mathbf{q},\mathbf{G}) = \langle \psi_{\mathbf{k}i} | e^{i(\mathbf{q}+\mathbf{G})\cdot \mathbf{r}} | \psi_{\mathbf{k}-\mathbf{q}j} \rangle \qquad (14)$$

and $\varepsilon_{kj}$, $|\psi_{\mathbf{k}i}\rangle$ the Kohn-Sham energy eigenvalues and eigenfunctions. [33].

In order to determine the transport spectral function $\alpha_{tr}^2 F(\omega)$, which allows us to calculate the DC resistivity, we first introduce the electron-phonon (EP) matrix element, which describes the scattering of an electron at the Fermi surface from state $|\psi_{\mathbf{k}i}\rangle$ to state $\langle \psi_{\mathbf{k}+\mathbf{q}j}|$ via the phonon perturbation:

$$\boldsymbol{\varphi}_{\mathbf{q}\nu} \cdot dV_{\mathbf{q}}^{eff}, \qquad (15)$$

where $dV_{\mathbf{q}}^{eff}$ is the change in the effective self consistent Kohn-Sham potential with respect to atomic displacements. The EP matrix element is then[34]:

$$g_{\mathbf{k}+\mathbf{q}j,\mathbf{k}i}^{\mathbf{q}\nu} = \sqrt{\frac{\hbar}{2M\omega_{\mathbf{q}\nu}}}\langle \psi_{\mathbf{k}+\mathbf{q}j} | \boldsymbol{\varphi}_{\mathbf{q}\nu} \cdot dV_{\mathbf{q}}^{eff} | \psi_{\mathbf{k}i} \rangle, \qquad (16)$$

where the phonon eigenvector $\boldsymbol{\varphi}$ of branch $\nu$ and momentum $\mathbf{q}$ is a solution to the dynamical matrix

$$\mathbf{D_q}\boldsymbol{\varphi} = \omega_{\mathbf{q}\nu}\boldsymbol{\varphi}_{\mathbf{q}\nu}, \qquad (17)$$

and $M$ and $\omega_{\mathbf{q}\nu}$ are the atomic mass and the phonon frequency respectively. The dynamical matrix was solved using Density Functional Perturbation Theory as implemented in the density functional code ABINIT [31, 32] with Fritz Haber Institute (FHI) pseudopotentials [35].

From the electron phonon matrix elements (16) we can now write down the spectral function:

$$\alpha_{out(in)}^2 F(\omega) = \frac{1}{N_F} \sum_{\nu} \sum_{\mathbf{k}i\mathbf{k}+\mathbf{q}j} \left| g_{\mathbf{k}+\mathbf{q}j,\mathbf{k}i}^{\mathbf{q}\nu} \right|^2 \eta_{out(in)} \\ \times \delta(E_{\mathbf{k}i} - E_F)\delta(E_{\mathbf{k}+\mathbf{q}j} - E_F)\delta(\omega - \omega_{\mathbf{q}\nu}) \qquad (18)$$

where the scattering efficiency factor $\eta$ accounts for the various scattering angles.

$$\eta_{out(in)} = \frac{\mathbf{v}_{\mathbf{k}i} \cdot \mathbf{v}_{\mathbf{k}(+\mathbf{q})j}}{\langle \mathbf{v}^2 \rangle} \qquad (19)$$

The transport spectral function is simply the difference between the integrated number of electrons scattered into or out of all states by all possible phonon modes:

$$\alpha_{tr}^2 F(\omega) = \alpha_{out}^2 F(\omega) - \alpha_{in}^2 F(\omega) \qquad (20)$$

From this, Allen[36] calculates the upper bound of the phonon limited DC resistivity by deriving a first order solution to the Boltzmann equation:

$$\rho_{DC}(T) = \frac{\Omega}{N_F \langle \mathbf{v}^2 \rangle} \int_0^\infty \frac{\alpha_{tr}^2 F(\omega)}{\omega} \frac{x^2}{\sinh^2 x} d\omega, \qquad (21)$$

where $x$ is $\omega/2k_B T$. From the DC resistivity we can approximate the temperature dependent Drude phenomenological damping term[29] via Equation (10)
Dynamical calculations were performed using the local density approximation to the exchange-correlation energy by Perdew and Wang [37] with a plane wave cutoff of 28 Ha. We use an unshifted $\mathbf{k}$-space grid of 16x16x16 resulting in 4096 points in the Brillouin zone, and 35 $\mathbf{q}$-points for the calculation of the dynamical matrix, corresponding to an 8x8x8 grid. Equilibrium lattice constants are given in Table 2. Optical calculations were performed using ground state wavefunctions from ABINIT on a 48x48x48 **k**-space grid, and the matrix elements (14) were evaluated using YAMBO [31] in a scheme that allows for the decoupling of the frequency and state dependence (see [38]).

It has been shown that quasiparticle corrections are necessary to accurately reproduce the band edge of silver [39]. DFT-LDA underestimates the optical gap $\omega_g$ which in silver leads to an overlap of the band edge $\epsilon''(\sim \omega_g)$ at the bulk plasmon frequency $\omega_s$. The additional loss contribution adversely affects the quality of any resonance near this frequency. In the case of silver both the Electron Energy Loss (EEL) plasmon peak and the bare film surface plasmon $\epsilon'(\omega) \approx -1$ have reduced quality compared to experiment. In Section IV we discuss the quality of materials in terms of the optical gap to screened plasma frequency ratio $\omega_g/\omega_s$. As DFT LDA tends to underestimate the optical gap, we expect an increase in the plasmonic quality of the alkali-nobles upon the inclusion of quasiparticle corrections.

## IV. RESULTS

The interband contribution to the imaginary permittivity is presented in Figure 1 and optical data for elemental silver and gold as well as the alkali-noble intermetallics is presented in Table 1. For comparison, we calculate the upper bounded $\rho_{DC}$ of gold and silver to be 2.66 μΩcm and 1.77 μΩcm respectively, in good agreement with the experimental values [29] of 2.04 μΩcm and 1.55 μΩcm. LiAg and LiAu have the lowest DC resistivities of 7.6 μΩcm and 8.9 μΩcm respectively, followed by NaAu and NaAg with values of 25 μΩcm and 33 μΩcm respectively. KAu has a resistivity of 70 μΩcm and for KAg the resistivity is the highest of the alkali noble intermetallics at 117 μΩcm. As the damping relies on the effective electron mass, the actual scattering rate is lowest for KAu, which has a bare plasma frequency of 1.54 eV. Both LiAg and LiAu have damping frequencies only 2 to 3 times greater than gold and silver, and reasonably high plasma frequencies of 7.28 eV and 7.2 eV respectively, resulting in $\gamma/\omega_p$ ratios which are comparable to gold.

Unfortunately, the optical gap in these materials is very small (0.12 eV and 0.33 eV respectively). This low band edge to plasma frequency ratio means that all surface plasmons will be screened by interband transitions, causing a reduction in the quality of high frequency/low permittivity phenomena.

The optical gap to plasma frequency ratio ($\omega_g/\omega_p$) gives a reasonable indication of the effect that interband transitions will have on a plasmonic material, however, it is probably more useful to describe the ratio in terms of the low energy dressed or screened plasma frequency $\omega_s$, which takes into account not only the energy of the optical gap, $\omega_g$, but also the magnitude and spread of the band edge. $\omega_g/\omega_s$ amounts to a description of the quality of the low permittivity behavior of a material. In terms of $\omega_g/\omega_s$, LiAg and LiAu do not perform as well as NaAu and KAu, and the damping frequency is too high in NaAg and KAg for them to perform well. In KAu, there is a small region of negative $\epsilon'$ above the plasma frequency due to the effect of the 2.8 eV transition. The real part of the dielectric function including the intraband contribution is shown in Figure 2.

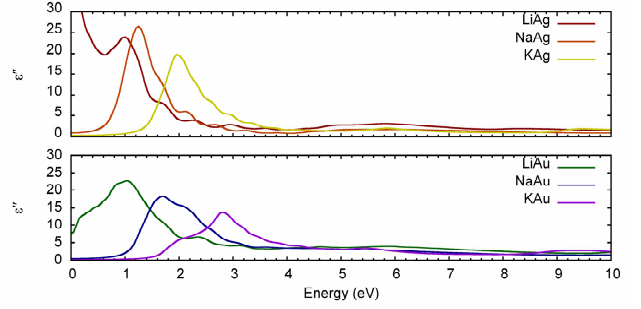

**Figure 1:** The interband contribution to the imaginary part of the permittivity for the alkali-noble intermetallics, including local field effects. Note that the optical gap for LiAg ($\omega_g$=0.12 eV) is less than the smearing width used in the approximation of the delta function in the denominator of Equation 13.

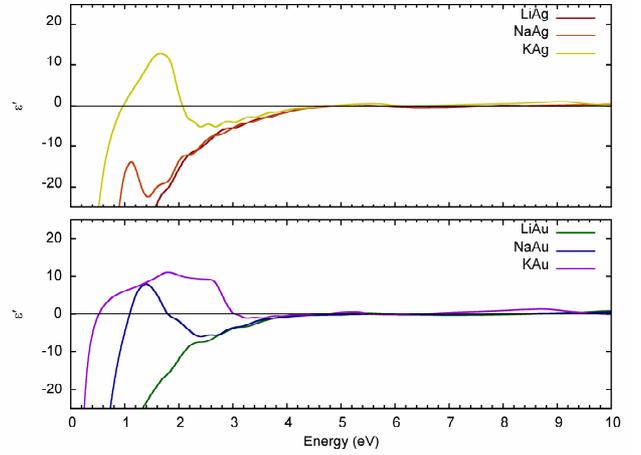

**Figure 2:** The real part of the permittivity including the intraband contribution for the alkali-noble intermetallics including local field effects.

**Table 1:** Optical and Electronic data for the alkali noble intermetallics. Temperature dependent quantities are given at 300K.

| Metal | $\omega_p$ [eV] [40] | Optical Gap $\omega_g$ [eV] | $\rho_{DC}$ [μΩ cm] | $\gamma$ [eV] | $\gamma/\omega_p$ | $\omega_g/\omega_s$ |
|---|---|---|---|---|---|---|
| Ag | 9.6 | 3.9 [41] | 1.77 | 0.0220 | 0.00229 | 1.0263 |
| Au | 8.55 | 2.25 [41] | 2.66 | 0.0262 | 0.00306 | 0.3879 |
| LiAg | 7.28 | 0.12 | 7.579 | 0.0540 | 0.00742 | 0.0254 |
| LiAu | 7.20 | 0.33 | 8.903 | 0.0621 | 0.00862 | 0.0742 |
| NaAg | 6.47 | 0.70 | 33.12 | 0.187 | 0.0288 | 0.0874 |
| NaAu | 4.84 | 1.13 | 25.45 | 0.0802 | 0.0169 | 1.0367 |
| KAg | 3.10 | 1.35 | 116..5 | 0.151 | 0.0486 | 1.387 |
| KAu | 1.54 | 1.55 | 70.07 | 0.0224 | 0.0145 | 2.9808 |

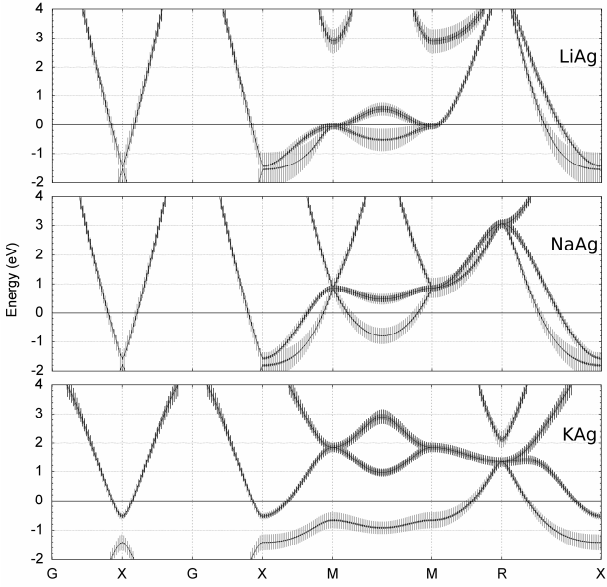

**Figure 3:** Band character of the alkali-Ag intermetallics. Width of gray bands is proportional to p-character, width of black bands proportional to s-character.

Due to the proximity of the band edge to plasma frequency, NaAu, KAg and KAu all have a finite region of positive $\epsilon'$ above the plasma frequency. This effect is also seen in elemental silver with a band edge to screened plasma frequency ratio above 1. It results in two additional bulk plasmons below the plasma frequency (excluding KAu). The higher energy plasmon decays into electron-hole pairs due to the over lap in frequency with interband transitions. The low frequency plasmon, however, is separated in frequency form the interband transitions, and as such should exhibit a strong EEL peak ($1/\epsilon''$). The main EEL peak in elemental silver reaches 1.43 [42], slightly smaller than in KAu, where it is 1.7, but larger than KAg and NaAu where it is 0.2 and 0.4 respectively.

The screened plasma frequency of NaAg is 8.01 eV, which is above the bare plasma frequency of 6.47 eV. This is due to a combintation of the magnitude of the band edge, its dispersion $d\epsilon''_{ib}/d\omega$, and its position with respect to the bare plasma frequency $\omega_g/\omega_p = 0.11$.

The frequency spread of the transitions in the alkali-gold compounds is due to the energy difference between the band pairs along R-X and X-M. Transitions that occur near X are responsible for the higher energy peaks comprising the band edge in LiAu (0.85 eV), but in NaAu and KAu, they comprise the lower energy peak (1.3 eV and 1.55 eV respectively).

The band structure and angular momentum character of the alkali noble intermetallics is presented in Figures 3 and 4. This 'fat-band' analysis has been calculated on a 16 x 16 x 16 k-grid within the Full Potential Linearised Augmented Plane Wave (FP-LAPW) code *elk* [43], using the all electron lattice constants from Table 2 and a smearing width of 0.001 Ha.

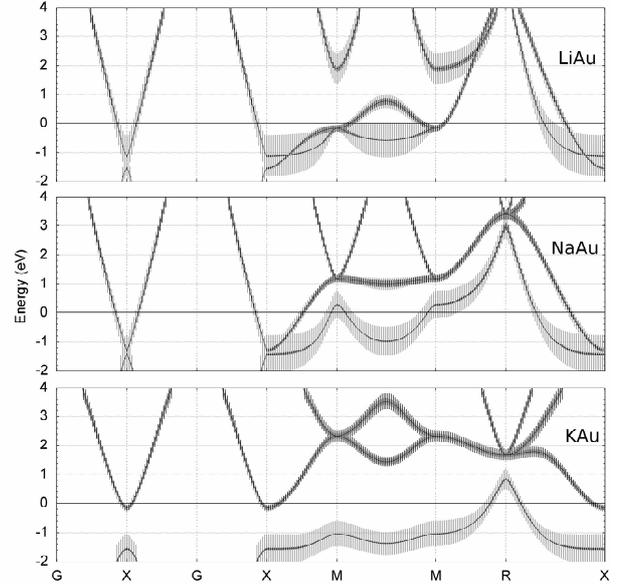

**Figure 4:** Band character of the alkali-Au intermetallics. Width of gray bands is proportional to p-character, width of black bands proportional to s-character.

**Table 2:** Comparison of All Electron (AE) and FHI Pseudopotentials (PP) Equilibrium Lattice Constants (LC). Calculated using *ABINIT* and *elk* respectively.

| Intermetallic | PP LDA LC [Bohr] | AE LDA LC [Bohr] |
|---|---|---|
| LiAg | 5.797 | 5.744 |
| LiAu | 5.688 | 5.714 |
| NaAg | 6.447 | 6.373 |
| NaAu | 6.324 | 6.308 |
| KAg | 7.083 | 7.210 |
| KAu | 6.891 | 7.071 |

In LiAu, the low energy transitions are primarily caused by transitions along the line M-M (from 0.32 eV), whereas the M-M midpoint is responsible for the higher energy transitions in NaAu and KAu (1.95 eV and 3.04 eV respectively). The 2.1 eV transition visible in LiAu occurs directly at M and is forbidden in the other gold compounds.

In LiAg there are two very distinct peaks, one at 0.12 eV, arising from the band edge along M-M, and a second due to transitions between parallel bands along a very brief segment of R-X. In NaAg and KAg these two sets of transitions overlap in energy and are difficult to distinguish.

The spreading of bands and reasonably high $\omega_g/\omega_p$ causes the screened bulk plasmon to be clear of interband transitions, resulting in excellent low permittivity behavior. $Q_{LSP}$ data for the Alkali-Noble intermetallics is shown in Figure 5. As predicted by the $\omega_g/\omega_s$ ratio, KAu has exceptionally good low permittivity $Q_{LSP}$, similar to that of silver, and better than gold.

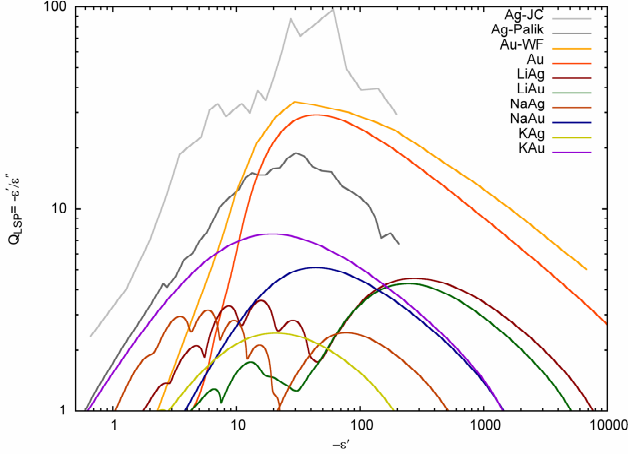

**Figure 5:** The resonance quality of a localized surface plasmon as a function of the real permittivity from first principles. Elemental optical data from JC: Johnson and Christy [44], Palik: E.D. Palik [42] and WF: Weaver and Frederikse [41].

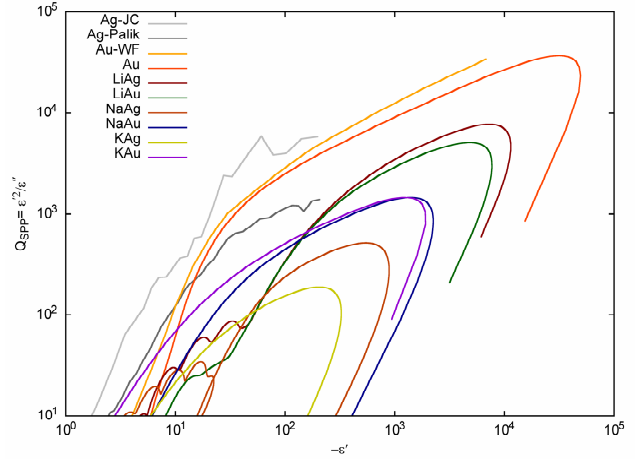

**Figure 6:** The resonance quality of a surface plasmon polariton as a function of the real permittivity from first principles. Elemental optical data from JC: Johnson and Christy [44], Palik: E.D. Palik [42] and WF: Weaver and Frederikse [41].

**Table 3:** $Q_{LSP}$ data for the alkali-noble intermetallics and gold calculated using Equations (22) and (23). We assume $\epsilon_\infty =1$. Equation (22) is used to approximate the frequency of maximum $Q_{LSP}$, we then use the interband contribution to the imaginary permittivity at this frequency to recalculate $Q_{LSP}$ using Equation (23)

| Metal | $\epsilon'_{ib}(0)$ | $\omega^{Q_{LSP}}_{max}$ [eV] Eqn (22) | $\epsilon''_{ib}(\omega^{Q_{LSP}}_{max})$ Eqn (22) | $Q^{max}_{LSP}(\omega^{Q_{LSP}}_{max})$ Eqn (23) |
|---|---|---|---|---|
| Au | 9.78 | 1.50 | 0.369 | 53.6 |
| LiAg | 52.3 | 0.560 | 20.1 | 5.97 |
| LiAu | 34.2 | 0.686 | 18.5 | 6.33 |
| NaAg | 14.1 | 0.928 | 10.9 | 3.16 |
| NaAu | 11.9 | 0.761 | 1.03 | 6.25 |
| KAg | 8.33 | 0.541 | 0.402 | 2.47 |
| KAu | 7.28 | 0.297 | 0.203 | 7.32 |

However, with a plasma frequency of only 1.54 eV, the frequency of maximum $Q_{LSP}$, $\omega^{Q_{LSP}}_{max}$ is 0.28 eV or 4.4 μm, and at $\epsilon'(\omega)=-1$, $\omega = 0.49$ eV, limiting its practical application.

The maximum $Q_{LSP}$ for NaAg occurs at quite high energies (3 eV), and is caused by contributions from lower energy interband transitions. In terms of plasmonic performance in the visible regime, NaAg performs best, as it has a $Q_{LSP}$ of 3.15 at 2.95 eV, which is much better than gold at this frequency ($Q_{LSP}$=0.16), and comparable to silver ($Q_{LSP}$=5.57).

There is a clear variation between the value of the Drude $Q_{LSP}$ maximum in Equation 6 and values obtained from the Drude intraband tail in Figure 3. This stems from the interband contribution to the real part of the permittivity. Although the ratio $\omega_g/\omega_s$ informs that the surface mode will not scatter into electron-hole pairs, the overall resonance quality is affected by the reduction in magnitude of $\epsilon'$ from interband contributions. This screening can be accounted for by replacing $\epsilon_\infty$ with $\epsilon_\infty + \epsilon'_{ib}(0)$, representing the zero frequency contribution to the interband component of the real part of the permittivity. This causes a decrease in $-\epsilon'$ below $\omega_g$ resulting in a decrease in low frequency quality. Additionally, for materials with $\omega_g < \omega_s$ the imaginary part of the interband component contributes to the reduction of $Q_{LSP}$ as well. A revised expression for the frequency of maximum $Q_{LSP}$ can now be derived from the addition of $\epsilon'_{ib}(0)$ to the real part.

$$\omega^{Q_{LSP}}_{max} = \left( \frac{\omega_p^2 - \gamma^2[\epsilon_\infty + \epsilon'_{ib}(0)]}{3[\epsilon_\infty + \epsilon'_{ib}(0)]} \right)^{1/2} \quad (22)$$

And the associated quality factor including contributions from $\epsilon''_{ib}(\omega^{Q_{LSP}}_{max})$ is approximately:

$$Q^{max}_{LSP}(\omega^{Q_{LSP}}_{max}) \approx \frac{2(\omega_p^2 - \gamma^2[\epsilon_\infty + \epsilon'_{ib}(0)])^{3/2}}{\epsilon''_{ib}(\omega^{Q_{LSP}}_{max}) + 3\gamma\omega_p^2 \sqrt{3[\epsilon_\infty + \epsilon'_{ib}(0)]}} \quad (23)$$

The effect of interband transitions is most noticeable in compounds with low $\omega_g$. In LiAg and LiAu, large permittivity behaviour is disrupted significantly by interband transitions. Both have damping to plasma frequency ratios only twice that of gold and as such they should have max $Q_{LSP}$ of 51.9 and 44.6 respectively. Table 3 shows the values of $Q_{LSP}$ calculated using equations (22) and (23). The actual values of $Q_{LSP}$ for LiAg and LiAu are approximately 10 times lower than they would be in the absence of interband transitions. The case is less severe for NaAg, NaAu and KAu, all approximately 4 times lower. For LiAg, LiAu and NaAg the dominant local surface plasmon damping mechanism at $\omega^{Q_{LSP}}_{max}$ is due to decay of the mode into electron-hole pairs, whereas NaAu, KAg and KAu—due to high $\omega_g/\omega_s$ ratios—escape this fate. Instead, screening by $\epsilon'_{ib}$ is the dominant damping mechanism.

Table 4: The effect of the inclusion of local fields on the magnitude of the band edge. Note that the band edge in LiAg is less than half the optical smearing width (0.1 eV) away from zero frequency.

| Compound | With Local Fields | | Neglecting Local Fields | |
|---|---|---|---|---|
| | $\epsilon''_{max}$ | $\omega_{\epsilon''_{max}}$ [eV] | $\epsilon''_{max}$ | $\omega_{\epsilon''_{max}}$ [eV] |
| LiAg | 62.8 | 0.00 | 63.3 | 0.00 |
| LiAu | 22.8 | 1.05 | 24.8 | 1.00 |
| NaAg | 26.6 | 1.25 | 30.6 | 1.20 |
| NaAu | 18.3 | 1.71 | 23.3 | 1.66 |
| KAg | 19.7 | 1.96 | 30.4 | 1.91 |
| KAu | 13.7 | 2.81 | 21.2 | 2.71 |

We now discuss the effect of interband transitions on propagating modes. The quality factor for propagating surface plasmons (Eqn (7)) for the alkali-noble intermetallics is presented in Figure 6. As this metric relies on the real part of the permittivity squared, materials with higher plasma frequencies perform substantially better than those with lower values. The maximum value of $Q_{SPP}$ is predicted within a factor of 2 by Eqn (7), the ratio $\gamma/\omega_p$ provides most of the pertinent information. In the absence of interband transitions, $Q_{SPP}$ in the Drude model is the upper half of an ellipse centered on the point $(\omega_p^2/2\gamma^2, 0) = (-\epsilon', Q_{SPP})$ with height $Q_{SPP}^{max} = \omega_p^2/2\gamma^2$. As a result, a majority of the increase in $Q_{SPP}$ with increasing $-\epsilon'$ occurs at lower permittivities. As interband transitions are introduced, disruption of low permittivity behavior causes the rate of increase $dQ_{SPP}/d(-\epsilon')$ to increase.

NaAg performs least favorably, with a high damping to plasma frequency ratio of 0.0288 causing a reduction in $Q_{SPP}^{max}$. The loop at $\epsilon'=-15$ is a result of the 1.15 eV peak in the real part of the spectrum. KAu and NaAu have similar performance at large permittivities, with similar $\gamma/\omega_p$ ratios of 0.0145 and 0.0169 respectively. KAu has slightly higher $Q_{SPP}$ at low permittivities due to the aforementioned effect of higher $\omega_g/\omega_s$. The effect of interband transitions is not visible in either spectrum. LiAg and LiAu have the highest $\gamma/\omega_p$ ratios and as such of the alkali-nobles they perform best, albeit at impractical permittivities.

## V. LOCAL FIELD EFFECTS

The difference in the maximum value of the interband contribution to the imaginary permittivity ($\epsilon''_{max}$) for the alkali-noble intermetallics with and without local field effects is shown in Table 4. Neglecting local fields amounts to taking the inverse of the $(\mathbf{G},\mathbf{G}')=(0,0)$ element of the microscopic dielectric function and making the assumption that the microscopic electric field varies slowly over the unit cell and has no effect on the field induced by the incident photon. Inclusion of local fields affects the optical properties of materials with localized states much more significantly than those without [45]. As the ionic character of the alkali-noble compounds increases from LiAg to KAu, the difference between $\epsilon''_{max}$ with and without local fields increases. The difference is negligible in the lithium compounds.

In the sodium compounds, the neglect of local fields increases the maximum interband transition magnitude by about 20%, and in the potassium compounds this is exacerbated to 50%.

## VI. CONCLUSIONS

We have discussed the optical properties of a number of alkali-noble intermetallics in terms of some generic plasmonic system quality factors. We have discussed the necessary electronic features for an ideal plasmonic material, and made adjustments to generic quality factors to account for the properties of real systems.

Most notably, we have calculated from first principles the Drude phenomenological damping frequency, allowing us to accurately describe the long-wavelength behavior of these materials.

Of the intermetallics discussed here, KAu performs best, outperforming gold at low permittivities, and over a small permittivity range, equaling the $Q_{LSP}$ of silver. Unfortunately, a low plasma frequency reduces the applicability of KAu to modern plasmonic applications, where high plasmonic quality in the UV-Visible frequency regime is desired. NaAg performs better than gold and comparably to silver at wavelengths around 400 nm. However, the plasmonic quality in these materials is dwarfed by the low permittivity properties of the elemental alkali metals, with potassium metal having a $Q_{LSP}$ of greater than 10 at 3 eV.

It is worth noting that small variations in the equilibrium lattice constant between different exchange correlation functionals causes bands near the FS to change from conduction (GGA) to valence (LDA) states. This has an impact on the locale of transitions in the Brillouin zone. Calculations of the dielectric function of lithium at high pressure have shown that the band edge can be shifted upwards in energy dramatically from 3 eV to 6.5 eV at 40 GPa. Additionally, self-energy corrections to the Kohn-Sham quasiparticle energies increase the optical gap.

Shifting the position of the band edge by even a small amount, especially in KAu and NaAu, will dramatically improve the frequency properties of the plasmonic response.

**Acknowledgements**
This work was supported by the Australian Research Council, and the University of Technology, Sydney. Computing resources were provided by the Australian Centre for Advanced Computing and Communication (ac3) in New South Wales and the National Facility at the Australian Partnership for Advanced Computing (APAC).